\documentclass[lettersize,journal]{IEEEtran}
\RequirePackage[hyphens]{url}
\usepackage{cite}
\usepackage{amsmath,amssymb,amsfonts}
\usepackage{algorithm}
\usepackage{algpseudocode}
\usepackage{graphicx}
\usepackage{textcomp}
\usepackage{xcolor}
\usepackage{etoolbox}
\usepackage{array}
\usepackage{xstring}
\usepackage[inline]{enumitem}
\usepackage{mathtools}
\usepackage{xspace}
\usepackage{url}
\usepackage{hhline}
\usepackage{multirow}
\usepackage{tabularx, booktabs}
\usepackage{pifont}
\usepackage{soul}
\usepackage{subcaption}
\usepackage[symbol]{footmisc}

\usepackage{ifthen}

\newboolean{showcomments}
\setboolean{showcomments}{true}
\ifthenelse{\boolean{showcomments}}
{ \newcommand{\mynote}[3]{
    \protect\fbox{\sffamily\scriptsize#1}
    {\small$\blacktriangleright$\textsf{\emph{\color{#3}{#2}}}$\blacktriangleleft$}}}
{ \newcommand{\mynote}[3]{}}

\definecolor{verylightgray}{gray}{0.8}
\definecolor{__red}{rgb}{0.8,0.1,0.1}
\definecolor{__green}{rgb}{0.1,0.6,0.1}

\usepackage{tikz}

\newcommand{\eg}{\textit{e.g.}\@\xspace}
\newcommand{\ie}{\textit{i.e.}\@\xspace}

\newcolumntype{G}{>{\centering\let\newline\\\arraybackslash\hspace{0pt}}m{0.12\linewidth} }
\newcolumntype{L}{>{\centering\arraybackslash} m{0.3\linewidth} }
\newcolumntype{M}{>{\centering\arraybackslash} m{0.15\linewidth} }
\newcolumntype{S}{>{\centering\arraybackslash} m{0.08\linewidth} }

\def\BibTeX{{\rm B\kern-.05em{\sc i\kern-.025em b}\kern-.08em
		T\kern-.1667em\lower.7ex\hbox{E}\kern-.125emX}}

\begin{document}

\title{The Key to Deobfuscation is Pattern of Life, not Overcoming Encryption}


\author{
  \IEEEauthorblockN{Taylor Henderson\IEEEauthorrefmark{1},
  Eric Osterweil\IEEEauthorrefmark{2},
  Pavan Kumar Dinesh\IEEEauthorrefmark{3} and
  Robert Simon\IEEEauthorrefmark{4}} \\
  \IEEEauthorblockA{Department of Computer Science,
  George Mason University \\
  Email: \IEEEauthorrefmark{1}thender8@gmu.edu,
  \IEEEauthorrefmark{2}eoster@gmu.edu,
  \IEEEauthorrefmark{3}pdinesh@gmu.edu,
  \IEEEauthorrefmark{4}simon@gmu.edu}
  \thanks{ This material is based upon work supported by the U.S.  Department of Homeland Security under Grant Award Number 17STCIN00001-05-00.  }
}

\maketitle

\begin{abstract}

  Preserving privacy is an undeniable benefit to users online. However, this
    benefit (unfortunately) also extends to those who conduct cyber attacks and
    other types of malfeasance. In this work, we consider the scenario in which
    Privacy Preserving Technologies (PPTs) have been used to obfuscate users
    who are communicating online with ill intentions. We present a novel
    methodology that is effective at deobfuscating such sources by synthesizing
    measurements from key locations along protocol transaction paths. Our
    approach links online personas with their origin IP addresses based on a
    Pattern of Life (PoL) analysis, and is successful even when different PPTs
    are used. We show that, when monitoring in the correct places on the
    Internet, DNS over HTTPS (DoH) and DNS over TLS (DoT) can be deobfuscated
    with up to 100\% accuracy, when they are the only privacy-preserving
    technologies used. 
    Our evaluation used multiple simulated monitoring points and communications
    are sampled from an  actual multiyear-long social network message board to
    replay actual user behavior. Our evaluation compared plain old DNS, DoH,
    DoT, and VPN in order to quantify their relative privacy-preserving
    abilities and provide recommendations for where ideal monitoring vantage
    points would be in the Internet to achieve the best performance. To
    illustrate the utility of our methodology, we created a proof-of-concept
    cybersecurity analyst dashboard (with backend processing infrastructure)
    that uses a search engine interface to allow analysts to deobfuscate
    sources based on observed screen names and by providing packet captures
    from subsets of vantage points.

\end{abstract}


\section{Introduction}

Privacy concerns are a top-of-mind consideration for many Internet users.
This has led to a growing abundance of protocols and tools that use cryptography to provide confidentiality guarantees.
In 2014, the Internet Architecture Board (IAB) issued a statement urging that protocols move toward using confidential operation by \emph{default}~\cite{iab-stmt-conf}, and most World Wide Web traffic was transmitted without encryption or authentication.
Flash forward to today, and the Web almost exclusively uses the Hypertext Transfer Protocol Secure (HTTPS)~\cite{rfc9110}, which relies on Transport Layer Security (TLS)~\cite{rfc8446} for confidentiality protections.
However, while protocols like these protect the confidentiality of home users, corporate employees, students, and more, they also protect miscreants who are conducting ranges of undesirable actions.
For example, confidentiality protections are used to obscure the sources of data leaks, document exfiltration, and more.
As a recent incident illustrates~\cite{thrush2023airman}, many of these threats are orchestrated in broad view, in public forums.
Troubling behaviors, where the sources are often obscured,
include cyber-bullying, disclosure of confidential or private documents, and much more.
When actions like these are taken by an actor who is using 
anonymizing technologies,
cybersecurity analysts currently have little (or no) ability to deobfuscate the source(s) of troubling communications.

%
Technologies like Transport Layer Security (TLS)~\cite{rfc8446}, DNS over TLS (DoT)~\cite{rfc7858}, DNS over HTTPS (DoH)~\cite{rfc8484}, Virtual Private Networks (VPNs), and many more are designed and deployed to preserve users' privacy (we call these \emph{Privacy Preserving Technologies}, or \emph{PPTs}).
Unfortunately, while PPTs can be a significant benefit for most users, they have also become a panacea for malfeasance.
Cybersecurity analysts who are routinely tasked with the duty of tracing the origins of document exfiltrations and/or determining the identities of actors involved in Transnational Criminal Organizations (TCOs) are stymied by the source obfuscation of encryption's privacy guarantees.
In the face of encryption, how can our security protectors track the origins of disclosures of confidential, or classified, documents?

In this work we focus on the problem of anonymous users posting dangerous (or confidential) content in public.
This includes posting threats, cyber-bullying, exposing classified or sensitive information, or any other disconcerting communications made in public, under the veil of anonymity.
Such situations may pose threats to individuals, children, national security, and many other potential victims.
We make a foundational observation: when transactions are known/knowable, such as posts on message boards, the \emph{Pattern of Life (PoL)} of those transactions (\ie posting and interacting) obviate the need to overcome PPTs' encryption in order to deobfuscate the sources of communications.

%
In this work, we produce a general methodology that deconstructs complex Internet topologies into more granular subsets (called \emph{``Scopes''}), uses features of observable network traffic and PoL to overcome PPTs, and deobfuscate sources.
This methodology is designed to allow the ingestion of network traffic (which may include PPTs) and facilitate a search-engine-like interface to deobfuscate otherwise hidden communication sources.
We use an approach called Topological Data Analysis (TDA)~\cite{zomorodian2012topological,wasserman2018topological,chazal2021introduction} to transform discrete network traffic into a univariate time series.
We show that, for this class of problem (\ie obfuscated sources posting to visible message boards), deobfuscation of sources is not only possible but nearly certain, from specific network vantage points.
Our results show that with a normalized accuracy score (from 0.0 to 1.0), we are able to deobfuscate sources using DoT and DoH with scores of 1.0, when analyzed from specific network vantage points.
Our methodology presumes no access to software, servers, or cryptographic keys, and performs deobfuscation based solely on visible network traffic.
%
%
%
%
Our contributions are:
\begin{itemize}
\item Our methodology, which ingests raw packet captures from variable networks and deobfuscates sources.
\item A Proof-of-Concept (PoC) dashboard and processing infrastructure that presents a search-engine interface for cybersecurity analysts, which allows them to upload network captures and perform deobfuscation of sources, ranking results based on accuracy and recall statistics.
\end{itemize}

The remainder of this paper proceeds as follows:  In Section~\ref{sec:bg} we present some basic background on the Domain Name System (DNS).  
Next, in Section~\ref{sec:method}, we explain our methodology, our reference architecture, and our algorithm.  
We present the results of our evaluation in Section~\ref{sec:eval}.
Then, we describe our future work in Section~\ref{sec:future} and related work in Section~\ref{sec:relwork}.
Finally, we offer some conclusions in Section~\ref{sec:conc}.  

\section{DNS Background}\label{sec:bg}

The Domain Name System's (DNS')~\cite{mockapetris-sigcomm88} resolution is a necessary first step in almost all modern Internet transactions.
As a result, the idea of remaining anonymous through the use of DNS has received a lot of attention.  

\begin{figure*}
\begin{center}
  \begin{subfigure}{0.45\textwidth}
    \begin{center}
    \includegraphics[width=0.95\textwidth]{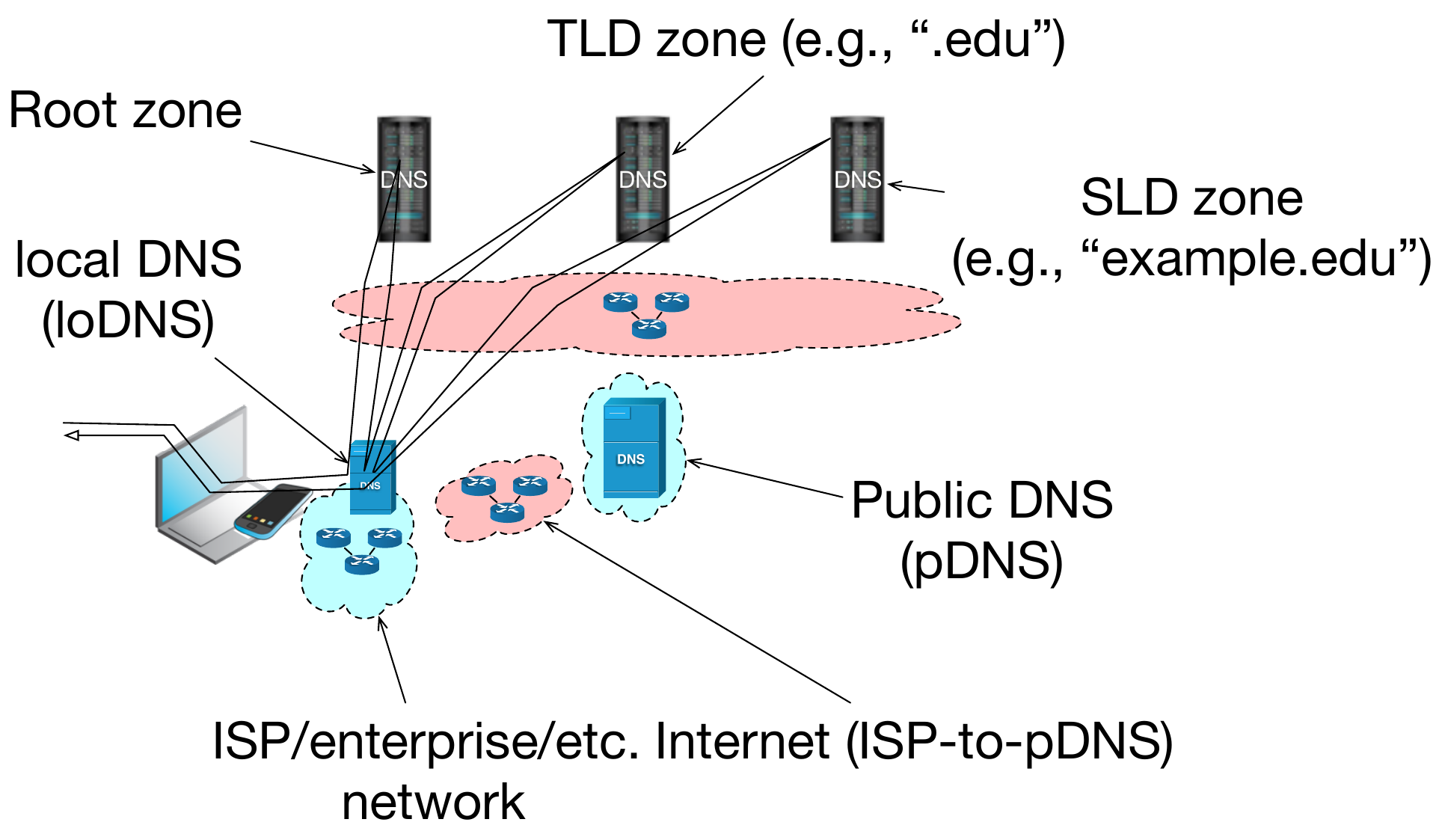}
      \caption{local DNS (loDNS)}
    \label{fig:lodns-config-usage}
    \end{center}
  \end{subfigure}%
  \begin{subfigure}{0.45\textwidth}
    \begin{center}
    \includegraphics[width=0.95\textwidth]{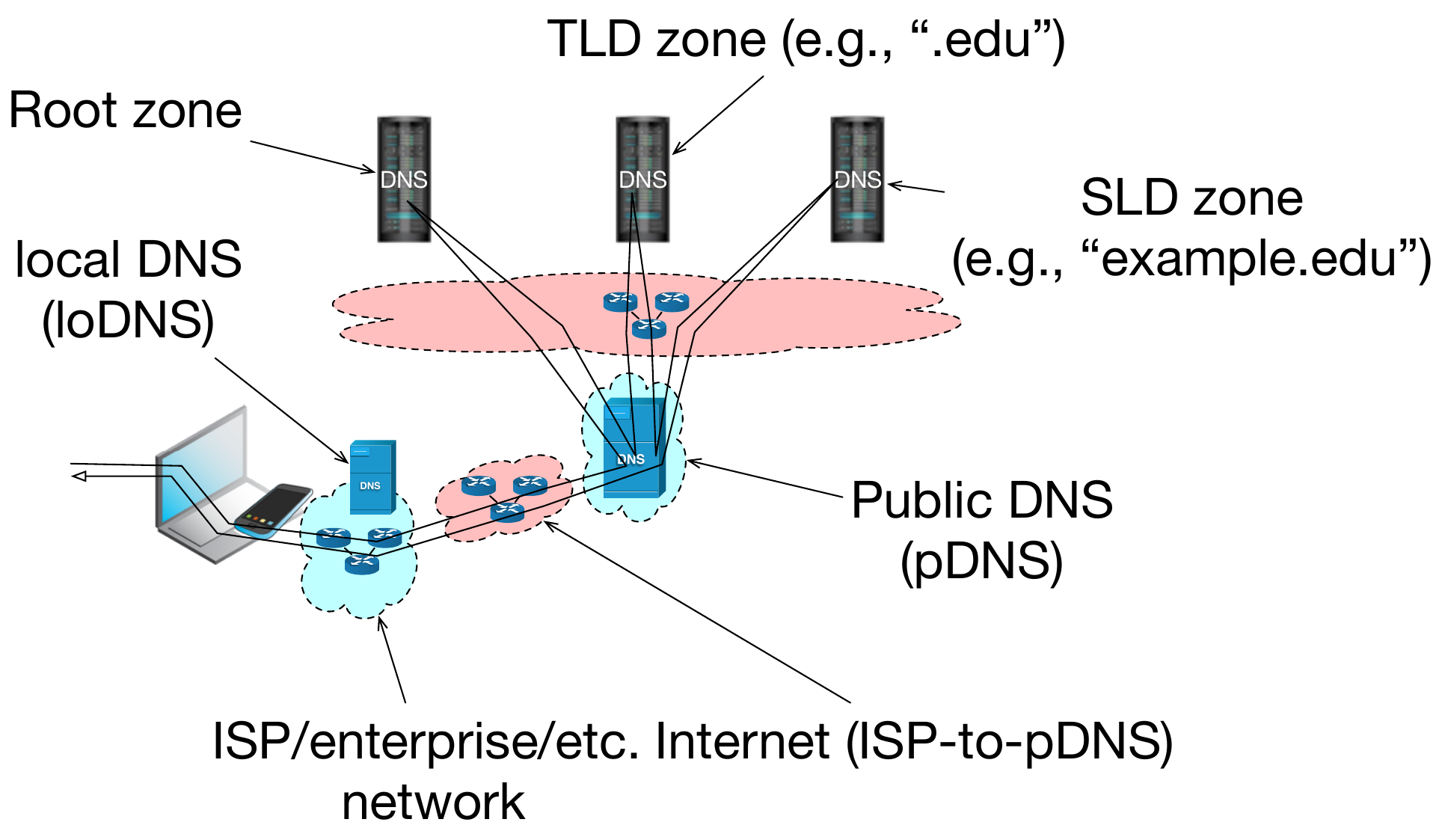}
      \caption{public DNS (pDNS)}
    \label{fig:pdns-config-usage}
    \end{center}
  \end{subfigure}
  \caption{This Figure contrasts using local DNS (loDNS) to public DNS (pDNS), with example traffic flows for each.}
  \label{fig:resolver-config}
\end{center}
\end{figure*}

%
%
Configuration changes are one way that users and operators have increasingly begun to bolster their security and privacy postures of DNS.
%
One
such configuration
change has been users' migration away from using local
DNS (``loDNS'') resolution infrastructures towards public DNS (``pDNS'') resolution.  
Typically, when users obtain IP addresses from DHCP, 
the host network (\eg,  an Internet Service Provider, a corporate enterprise, a guest WiFi network, etc.)
will also include the address of a loDNS resolver. This is so the system can begin
resolving DNS names as soon as the IP address is leased, as shown in Figure~\ref{fig:lodns-config-usage}.
%
This configuration is typically a convenience for the user, but also can be a security measure taken by the host
network operator.  Nevertheless, some users prefer to shift this to third party DNS operators, as illustrated in Figure~\ref{fig:pdns-config-usage}.
In 2009, Google announced its offering of Public DNS (a pDNS option)~\cite{honest-dns-announcement} and users began to consider the value proposition 
(illustrated in Figure~\ref{fig:resolver-config}) 
of
moving from loDNS to pDNS (now offered by many other large providers such as Cisco's OpenDNS, Cloudflare's {\tt 1.1.1.1}, Quad9's {\tt 9.9.9.9}, Neustar, and more).
In some cases, these choices are motivated by service offerings of pDNS providers, which can range from parental-controls that
help safeguard minors' web browsing, to malware connection prevention, and beyond.  In many cases, the motivation
has been stated as the pDNS provider will be better able to offer security and configuration assistance, regardless
of users' source networks.  

In addition to configuration changes, the protocols and system technologies themselves are also evolving to address privacy concerns.
In regards to protocol and system evolution,
the Internet Engineering Task Force (IETF) has taken up the challenge to
incorporate privacy concerns and protections into many new protocols and/or extensions to existing protocols.
A couple of general observations can loosely frame some of the motivating themes for these works:
1) user transaction information is sensitive and should be obscured by encryption where possible and 
2) meta-data about transactions, services, and end-points should be protected.


Many
different approaches have been (and are being) tried to enhance privacy in the DNS.
Some notable protocol extensions (though, there are others) that are going through experimentation, standardization, and deployment consideration are
establishing DNS over Transport Layer Security (TLS) connections (DoT)~\cite{rfc7858}
and DNS over HTTPS (DoH)~\cite{rfc8484}.

The DNS-over-TLS (DoT)~\cite{rfc7858} protocol is separated into two stages.  Initial work has explicitly focused on using TLS to protect
the transaction data between a stub-resolver and a recursive resolver during DNS resolution.  Many of the 
motivations for this work arose after users began to purposely use public DNS resolvers (pDNS).  
When using DNS-over-TLS, queries and responses are sent over TLS to port 853 (rather than DNS' canonical port, 53).
Then, the recursive resolver nominally uses the conventional DNS protocol to authoritative name servers.  An addition to 
DoT is being considered by the IETF, whereby resolvers would be able to use TLS to connect to authoritative
name servers.  This is being called Authoritative DoT (ADoT). 
With the addition of connection-oriented transactions (a result of using TCP), and with the complication of certificate
validation (necessary to complete the TLS handshake), and the increased processing overhead of TLS' encryption.

The DNS-over-HTTPS (DoH)~\cite{rfc8484} protocol proposes to allow client software to perform DNS
queries by wrapping them in HTTPS and sending them directly to DNS resolvers that support the DoH protocol (over HTTPS' canonical
port, 443).  While this may seem to be a direct analog of DoT, it differs in very substantial (though nuanced) ways.  In particular,
because the determination of where to send DNS queries is now being made \emph{above} the operating system layer, users and 
administrators have no ability to \emph{know} whether applications have chosen different resolvers.  DoH enables applications,
apps, and other software to establish channels that would not otherwise be (at least) detected.
DoH also
brings with it most (if not all) of the operational complexities, cryptographic overheads, and other implications 
of the DoT protocol.  


\section{Methodology}\label{sec:method}

Among our most foundational findings is that the inter-communication timing
pattern in which users interact online (such as times a user posts messages on
message boards) is an observable that is (itself) sufficient to deobfuscate and
perform attribution if measurements can be taken from a specific network
vantages with one second time resolution.
Our basic observation is that even if traffic is encrypted by a PPT, the set of postings (at specific times, from specific network locations) forms a distinguishable fingerprint of user activity.
This \emph{Pattern of Life (PoL)} (\ie, the temporal pattern of message postings, and related meta-transactions to DNS) is sufficiently distinguishable so that overcoming encryption is not necessary for deobfuscation.
To quantify this 
PoL, we use techniques from
statistical time series analysis.

Our next observation is that knowing \emph{where} to instrument monitoring is the next key (after PoL) for deobfuscation.
To simplify the Internet's topology, without loss of generality, we developed a projection of the Internet's large and complex structure into a reduced set of fixed stages of protocols' control flow, which we call \emph{``Scopes''}.


\subsection{Scopes}\label{sec:scopes}

The Internet is a large and complex system, composed of hundreds of thousands (if not millions) of separate administrative networks.  The Classless Inter-Domain Routing (CIDR) Report~\cite{cidr-report-2023} measures upward of 931,903 separately routable network prefixes that are distributed across over 74,728 distinct Autonomous Systems (ASes), as of May 2023~\cite{cidr-report-2023}.
In order to use our methodology to deobfuscate network sources that are using Privacy Preserving Technologies (PPTs), measurements must be made at specific locations along the network paths between clients and the services they transact with.  Because PPTs operate in very different ways from each other, our generalizable approach leverages knowledge of how these technologies specifically operate and 
coarsens
the Internet's complex technology.
From the starting premise that being able to observe all network paths throughout the Internet would be infeasible but would enable deobfuscation, we have constructed a methodology to identify what specific subset of locations are necessary and sufficient.

\begin{figure}
\centering
\includegraphics[width=0.45\textwidth]{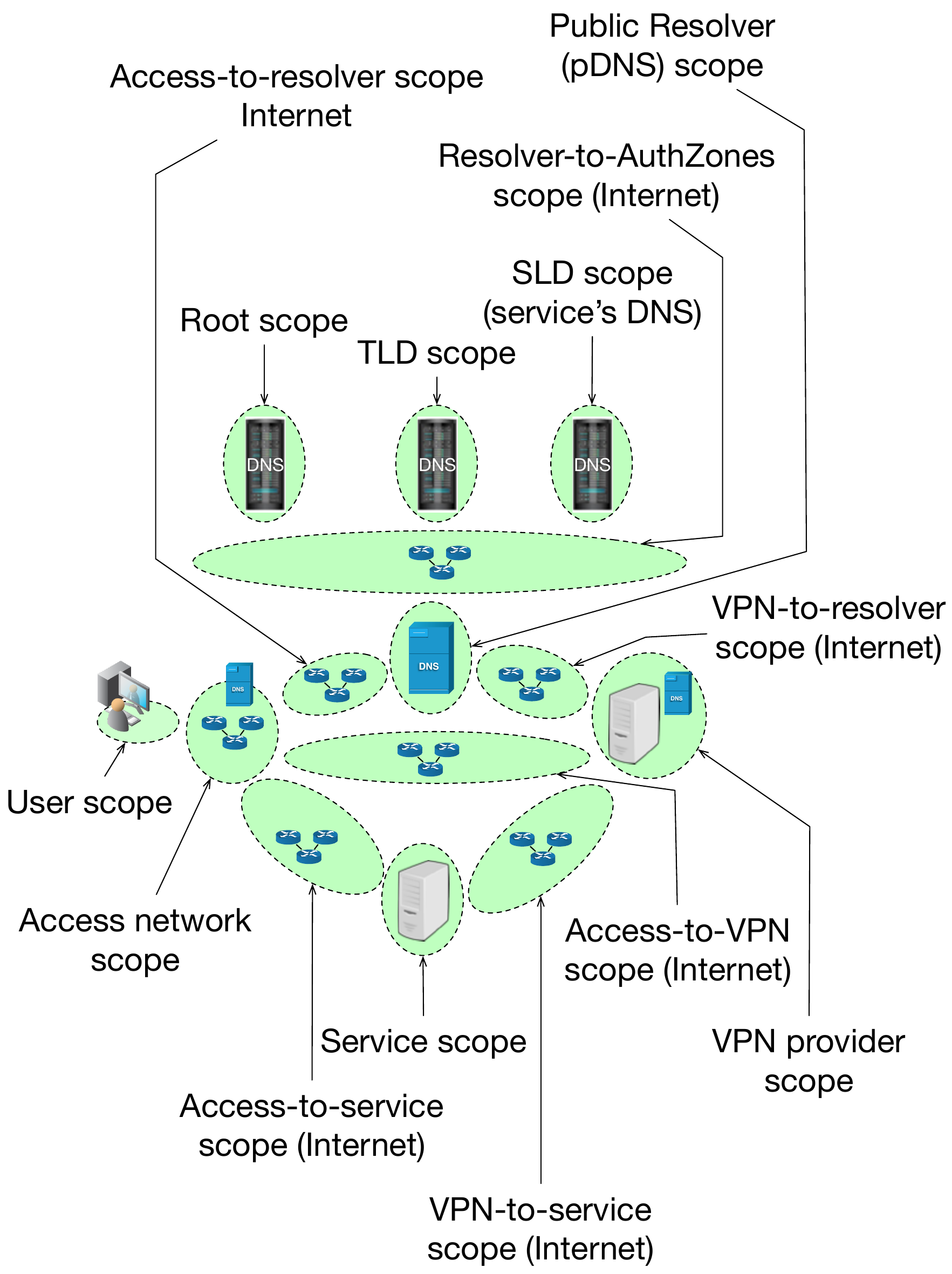}
\caption{This Figure illustrates the network scopes involved in user communications for DNS and with an Internet service, such as a message forum.}
\label{fig:scopes}
\end{figure}

The intuition behind scopes is to project the protocols' control flows that are being used.  In order to transact with an Internet service (such as a message forum), users first perform Domain Name System (DNS) resolution~\cite{mockapetris-sigcomm88}, followed by a transport-layer connection to a service (\eg a message forum service).  Each of these stages can take on different forms that depend on how service operators deploy their infrastructures.  Some web-based message forums may use Content Distribution Networks (CDNs), some may use local instances at fixed locations.  Similarly, some DNS providers offer resolution services from single networks, and some distribute across many locations.  
Furthermore, there are vast and complex topologies of inter-domain routing networks that often exist between endpoints of communication.  
In order to accommodate the varying degree of complexity in different networks, Scopes quantizes the protocols' communications into general regions (illustrated in Figure~\ref{fig:scopes}).
For example, when a user configures a public DNS resolver, and that service iterates through the DNS delegation hierarchy to resolve the domain name of the website for a message forum, we break this into the following Scopes:
\begin{itemize}
\item The user's Access network, or their Internet Service Provider (ISP) scope: where all of a user's traffic originates and returns to.
\item The Access-to-resolver scope: the set of inter-domain networks that connect the ISP to a public resolver.
\item The public DNS (pDNS) resolver scope: the network(s) responsible for operating the DNS resolution service.
\item The Resolver-to-Authoritative Zones scope: the potentially large portion of the Internet where inter-domain transit routing occurs.
\item The DNS root zone scope: the networks responsible for DNS resolution of the root zone.
\item The Top-Level Domain (TLD) scope: where resolution of a given TLD (such as {\tt .com}, {\tt .edu}, {\tt .org}, etc.).
\item The Second-Level Domain (SLD) scope: the specific authoritative zone for an Internet service (such as {\tt gmu.edu}, {\tt dhs.gov}, etc.).
\end{itemize}
Next, we use a similar projection method to model the control flow of a user transacting with an Internet service.  The scopes involved in these transactions are:
\begin{itemize}
\item The user's Access network (ISP), again.
\item The Access-to-service scope: an amalgamation of the large set of inter-domain networks that connect users' access/ISP networks to a specific service scope.
\item The Service scope: the specific network(s) that an Internet service (like a message forum) are operated on.
\end{itemize}

In addition to the above scopes that befit a basic Internet transaction (whether an HTTP~\cite{rfc9113}, HTTPS, TLS~\cite{rfc8446}, etc.), our Scopes methodology includes semantics for when other common PPTs are used.  
Some PPTs, such as DNS-over-TLS (DoT) and DNS-over-HTTPS (DoH), 
implement their protections along the same control path as plain old DNS (poDNS).  As a result, these PPTs make use of the same network scopes that are used for poDNS.
In contrast, however, some PPTs use different control flows.  For example, Virtual Private Networks (VPNs) create encrypted tunnels between endpoints.  One common form of VPN is for clients to encrypt network traffic and send it to a VPN provider, and for that provider to act as the source/destination of the resulting (unencrypted) network flows across the Internet.  This has the effect of making the VPN provider's network appear to be the endpoint of clients' Internet traffic, but results in tunneled communications between the provider and the client.
To accommodate this PPT, we are able to extend the existing Scopes to include:
\begin{itemize}
\item Access-to-VPN scope: where VPN-tunneled traffic is sent from a client's access/ISP network to the VPN provider.
\item The VPN Provider scope: where encrypted tunnels terminate, and where de-encapsulated client traffic is sent to Internet destinations (appearing to come from that VPN endpoint/egress).
\item The VPN-to-resolver scope: which is a similar amalgamation to the Access-to-resolver scope (defined above), but separated in our model as it may have some degree of disjointness from other transit paths.
\end{itemize}


Some of the scopes that are necessarily involved in communications are inherently composed of single (or a small number of) administrative organizations.  Examples include the ISP scope, to DNS root zone scope, each TLD scope, each SLD scope, VPN provider scopes, and the Service scopes.  
We classify these scopes are \emph{low-order scopes}, because of the low-order of administrative entities that they contain.
While the administrative diversity within these scopes is small, the Internet is composed of many instantiations of these low-order scopes.  
Conversely, some other scopes amalgamate large numbers of inter-domain ASes.  Examples of these include the Access-to-resolver scope, Resolver-to AuthZones scope, and the VPN-to-service scope.
We broadly classify these other scopes as \emph{amalgamation scopes}.  

\begin{figure*}
\begin{center}
  \begin{subfigure}{0.45\textwidth}
    \begin{center}
    \includegraphics[width=0.95\textwidth]{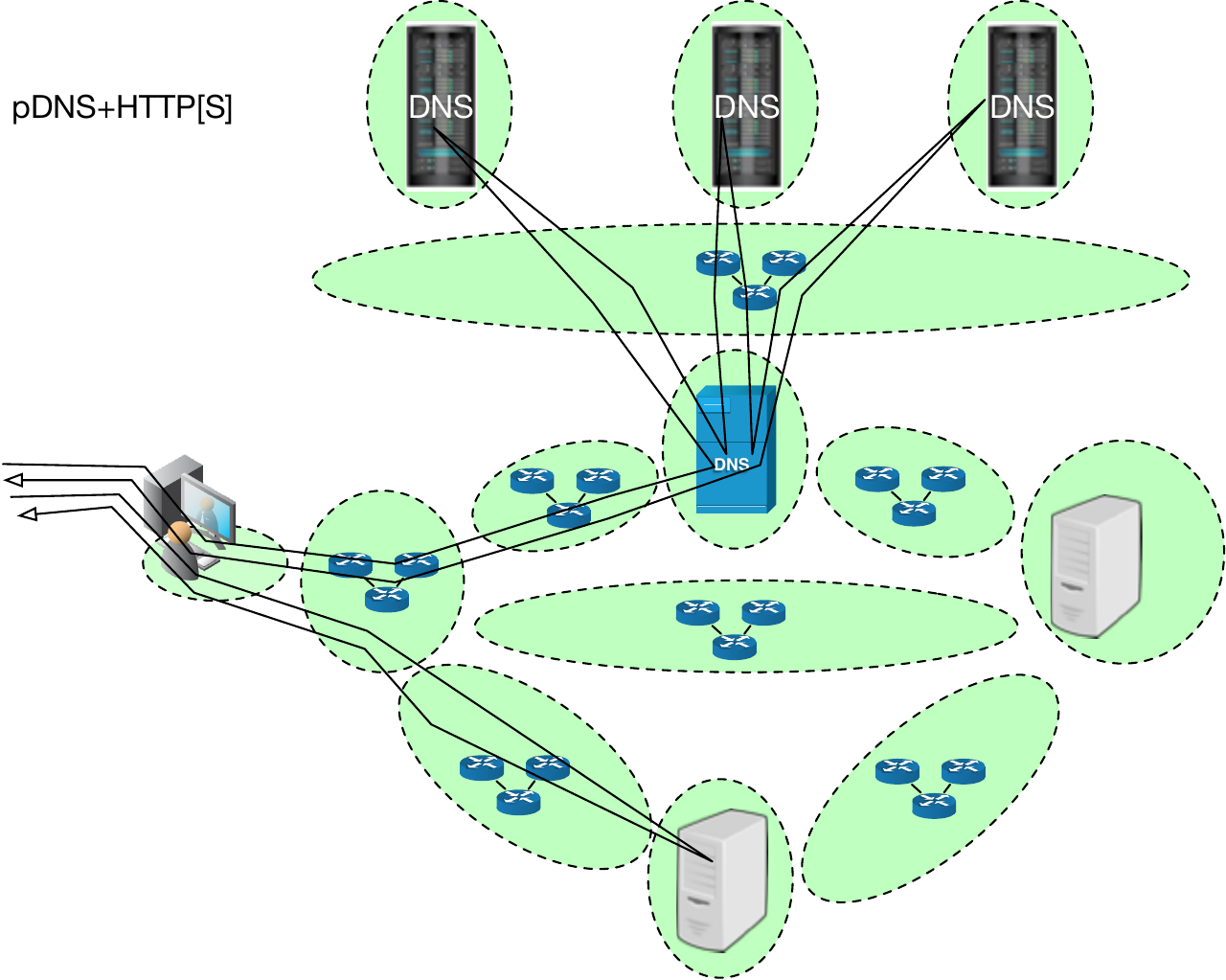}
      \caption{public DNS (pDNS)}
    \label{fig:ex-pdns}
    \end{center}
  \end{subfigure}%
  \begin{subfigure}{0.45\textwidth}
    \begin{center}
    \includegraphics[width=0.95\textwidth]{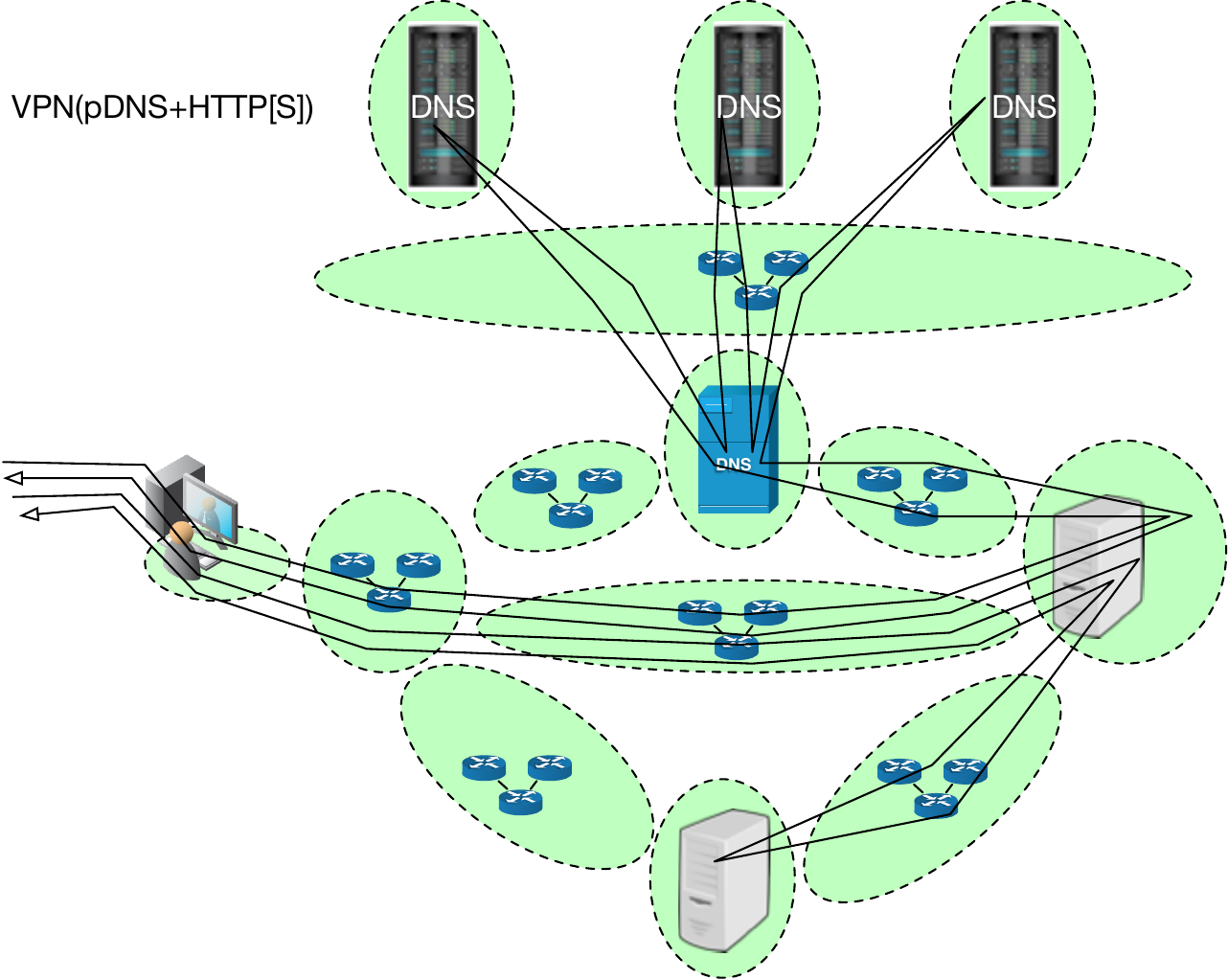}
      \caption{VPN}
    \label{fig:ex-vpn}
    \end{center}
  \end{subfigure}
  \end{center}
  \caption{This Figure illustrates the follow of data across scopes for public DNS (pDNS) and using a VPN.  The traffic flows cross the same scopes
  whether encryption is used or not (pDNS, DoT, DoH, etc.) are all encompassed.}
  \label{fig:proto-traces}
\end{figure*}

\vspace{4mm}
\noindent {\bf Example Control Flow:} 
When users make use of a public DNS resolver (pDNS), such as {\tt 1.1.1.1}, {\tt 8.8.8.8}, etc., Figure~\ref{fig:ex-pdns} illustrates how we map the resulting network traffic into our methodology.  Similarly, Figure~\ref{fig:ex-vpn} depicts
the control flow of protocol traffic when users employ a VPN with a pDNS resolver.  These figures illustrate which network scopes traffic traverses, and where relevant measurements can be made.  Depending on which set(s) of PPTs are being used, different segments of traffic may be encrypted on in the clear.  However, our foundational observation is the \emph{pattern of life} of these network communications is necessary for these protocols.

\vspace{4mm}
\noindent {\bf Deobfuscating with Scopes:} 
The Scopes methodology allows us to pinpoint which sets of scopes are necessary and sufficient to monitor in order to deobfuscate sources. 
Our results illustrate where high accuracy, high recall@k, and low rank are achievable from sets of \emph{low-order scopes}.  
These scopes are the most useful, compared to \emph{amalgamation scopes} which are not, because the latter include size and diversity that could make monitoring intractable.  
For example, results that indicate fulsome monitoring is necessary in the Resolver-to-AuthZones scope (an \emph{amalgamation scope}) would imply needing to monitor all resolver traffic throughout the inter-domain transit space of the Internet.  
Instead, we center our evaluation on \emph{low-order} scopes.  
Whereas gaining the cooperation of some of these may prove to be prohibitive, the feasibility of instrumenting them is far more realistic.


\subsection{Model Overview}\label{sec:model}
This subsection is provided to give an overview for the high level steps of the algorithms used.
Detailed pseudocode for the algorithm is provided in Appendix~\ref{app:algo} and in depth steps
describing why each step was included, the values used for each algorithm and the process taken
to arrive at each value is provided in the following subsections.

The model used for preprocessing the data and deobfuscating users to IP
addresses follows the following stages. The hypothesis behind this approach is
that when users communicate online in a live discussion, they leave traces of
their communication throughout the Internet. These traces are time-varying
signals, time series, which can identify the user. These time
series persist to varying degrees even when users attempt to hide their
identity with PPT. These identifying features' performance can be compared, and 
the degree of privacy per scope can be quantified.

\begin{enumerate}
    \item 	Data from PCAPs is grouped on a per IP basis: if the IP is seen in the packet or occurs within a time window of other packets being sent (Appendix Algorithm~\ref{algo:data} Lines 1--15)
    \item 	Feature selection based on available scopes (Appendix Algorithm~\ref{algo:data} Line 17)
    \item 	Sliding-sum over features extracted for each scope (Appendix Algorithm~\ref{algo:data} Line 18)
    \item 	Features are transformed using Topological Data Analysis-Persistence Landscape to capture multivariate PoL (Appendix Algorithm~\ref{algo:data} Line 19 and Appendix Algorithm~\ref{algo:tda})
    \item 	Above steps are repeated for features of the service of interest (\eg a message board) (Appendix Algorithm~\ref{algo:log})
    \item 	PoLs seen in scopes are compared in PoL of the users on the service of interest using normalized cross-correlation for the time the persona was using the service (Appendix Algorithm~\ref{algo:sim} and Algorithm~\ref{algo:link})
    \item 	Time series with the highest normalized cross-correlation are determined to be the same persona. (Appendix Algorithm~\ref{algo:link})
\end{enumerate}

\subsection{Data Preprocessing}

First the time series on a per user basis must be created from the raw data
available at the scopes. Our methodology is designed to ingest packet captures
from separate network scopes, $Scope \in \mathbf{Scopes}$. All scopes were
included initially for the experiments, and subsets of scopes and features were
evaluated to provide recommendations for deployment. The model accepts PCAPs,
although other formats that capture the same features can be used. Data is
grouped into data frames, tables/arrays used by machine learning algorithms, by
IP address. This grouping isolates different user's features before
transforming them into time series. Each IP address's POL also captures
information indirectly related to them (\eg, communications that could have
resulted from a client communication/other client communication that the client
could use, estimates of cached data). It is only possible to determine if one
packet is related to another with complete knowledge of each system involved.
As the goal of this model is to work without complete knowledge of every
system, it must make an educated guess if a packet is related to other users.
The experiments presented here use a simple model for this purpose and use
steps later on, TDA, to attempt to remove noise in the data. A small time
window is used to detect data that may be related to an IP address but does not
have the IP address in the packet. A packet is related to the IP address if a
packet from or to the IP was sent within the window of the other packet.
Equation~\ref{eq:dataframes} defines this set of packets per IP. This is to
best estimate the information that could influence a client's state. The time
window used is based on the DNS TTL for the domain/message board of interest.
This process is carried out for each scope individually because different
features may appear different and have varying degrees of important information
encoded. Pseudocode for this step can be found in Algorithm~\ref{algo:data}.

\begin{multline*} 
    ipHit_{ip} = \{packet \in \mathbf{Scope} \\
    | packet.src = ip \lor packet.dst = ip \}
\end{multline*}

\begin{multline*}
    cacheHit_{ip} = \{packet \in \mathbf{Scope} \\
        | \min(packet.time - p.time | p \in d_{ip}) < \mathbf{S}.cache.ttl\}\\ 
\end{multline*}
\begin{equation}
    d_{ip} = ipHit_{ip} \cap cacheHit_{ip} \label{eq:dataframes} 
\end{equation}

Feature selection is then performed on the dataframe to preserve only the most
performant features (\eg, IP packet length, per-flow inter-segment time
spacing). A complete list of features extracted and tested can be found in
Table~\ref{tab:features}. These features were selected because of their ability
to be transformed into time series and contain
state or personal information. Additionally, previous work has shown time and
packet length as key identifiers in machine-learning models for
deanonymization~\cite{deepcorr}~\cite{darkdetect}. 
Deployment of the model does not require all features from this list. Each
feature was tested independently and in combination with other features. Not every feature is available at every scope, and some PPTs block our
insight into some features. If this happens, the feature is removed from
evaluation at the given scope in these cases. This is done by detecting
features that have zero variance across all clients.

\begin{table}
    \begin{center}
        \begin{tabular}[c]{|l|l|}
            \hline
            {\bf Packet Field} & {\bf Description} \\ \hline
            \hline
            frame.time & Packet Timestamp\\
            ip.src & IP Source Address\\
            ip.dst & IP Destination Address\\
            ip.proto & Transport Protocol\\
            ip.len & Packet Size\\
            tcp.dstport & TCP Destination Port\\
            tcp.connection.syn & TCP SYN Flag State in Packet\\
            tcp.ack & TCP ACK Flag State in Packet\\
            tcp.len & TCP Segment Length\\
            tcp.reassembled.length & TCP Reassembled Len\\
            HTTP.request & HTTP Request Data\\
            tcp.time\_relative & Time Since Flow Began \\
            tcp.time\_delta & Inter-segment Time Delta\\
            tcp.payload & Data in TCP Segment\\
            dns.qry.name & DNS Query Name\\
            dns.opt.client.addr4 & EDNS0 Client Subnetv4 \\
            dns.opt.client.addr6 & EDNS0 Client Subnetv6 \\
            dns.opt.client.addr & EDNS0 Client Subnet\\
            \hline
        \end{tabular}
    \end{center}
    \caption{Features extracted from PCAPs.}
    \label{tab:features}
\end{table}

Based on the extracted features in each scope, the model transforms each
feature per scope per IP into a time series using a one-second sliding sum with
a one-second skip as seen in Equation~\ref{eq:slidingWindow}. This is done to
transform tabular data into a time series. A rolling sum was chosen to avoid
losing any variability in the communications. A one-second time window was
chosen through a grid search of possible values. These time series can be
subselected to form different univariate and multivariate time series.

If a multivariate time series is chosen, it must be transformed into a
univariate time series. This is done using the techniques from topological data
analysis in this evaluation. Specifically, the $L_2$ norm of the series's TDA persistence
landscape (TDA-PL) is based on the work done
in~\cite{gidea2018topological}. This method was chosen over others because it
has been shown to be able to capture the topological and geometric information of
multivariate time series for similar problems. Additionally, the output of the
persistence landscape function exists in a measure space under the $L_2$ norm, allowing it to be used with traditional machine learning algorithms and
techniques. 

The method takes an input multivariate time series, $t$, and splits it up using
a sliding window with user-defined size and skip, such as in
Equation~\ref{eq:plSliding}. This results in a series of point clouds. For each
of these windows, the Vietoris-Rips filtration, rips filtration, calculates the
persistent homology in the form of a set of birth-death pairs. Persistent
homology will compute topological features of a point cloud by building up a
simplicial complex of the point cloud and capturing how the topology changes
over time. A simplex is a generalization of a triangle to higher dimensions. A
simplicial complex is a collection of simplices where, for each n-simplex, all
(n-1)-simplices are included in the complex. The Vietoris-Rips filtration was
chosen due to its speed at computing a good representation of the persistent
homology. This is helpful due to the time complexity of the inherent mathching
problem. The Vietoris-Rips filtration computes a simplicial complex by ordering
all point pairs by their distance from each other. It then begins adding
0-simplices and 1-simplices (lines between points) to the simplicial complex in
order from closest points to further away points. It forms a higher dimensional
simplex for a group of points if they all have 1-simplices between them (they
are fully connected). The Vietoris-Rips filtration requires a maximum distance
for its computation where it will not stop adding 1-simplices when the distance
between the points involved is above this value. Additionally, this set can be
limited to simplices of a max dimension, known as a $n$-skelton. This is done
to focus the analysis on specific types of topological features. This is
desired as persistent homology is an expensive computation. When persistent
homology is computed it quantifies when a topological feature is first and last
seen and stores these values as the birth and death, respectively. These
birth-death pairs encode the persistent homology of the point cloud. The set of
birth-death pairs for each window can not be easily compared to each other and
are not stable under perturbation. Different functions can move the birth-death
pairs into a space where statistical machine-learning algorithms can be
applied. Our model uses the persistence landscape function to gain order to
gain the properties mentioned. This moves the birth-death pair sets into a
measure space. The persistence landscape takes each birth-death pair and turns
it into two connected line segments. For a given birth-death where the birth
with $b$ and the death $d$ the first line segment begins at $(b, 0)$ and ends
at $(((d-b)/2)+b, (d-b)/2)$. The other line segment for the pair starts at
$(((d-b)/2)+b, (d-b)/2)$ and ends at $(d, 0)$. This transformation is done for
all birth-death pairs in the set. The persistence landscape then defines the
$k^{th}$ landscape as the set of $k$-max points of the persistence landscape.
The resulting persistence landscapes can then be compared to each other using
the $L_2$ norm. The algorithm used for TDA can be found in Algorithm~\ref{algo:tda}. This
transforms the landscapes into a single-point summary of the persistent
homology. Putting these together into a time series gives insight into how the
topology of our original multivariate time series varied over time. This
process is summarized in Equation~\ref{eq:PL}. A grid search of possible
hyperparameters for TDA-PL was done to determine the best values. The grid
search was run over the window size, window skip, $n$-skelton/dimension of the
homology, number of persistence landscapes to keep, and Vietoris-Rips max
distance. The complete similarity function discussed here can be found in Algorithm~\ref{algo:sim}.

\begin{multline}
    ts_{ip} = x_0,\ldots,x_n \\
    x_i = \sum\{p \in features_{ip} | \\p.time \geq i \land p.time < i+1 \} \label{eq:slidingWindow}
\end{multline}

\begin{multline}
    W_{ip} = x_0,\ldots,x_n \\
    | x_i = y_i,\ldots y_{i + windowSize} | y_j = ts_{ip}[j]
    \label{eq:plSliding}
\end{multline}

\begin{multline}
    univariateTimeSeries_{ip} = \\
    \{ L_2(PersistenceLandscape(VietorisRips(w))) \\
    | w \in W_{ip}\}
    \label{eq:PL}
\end{multline}

The data of communications on the server can be captured in different manners, but
should be transformed so that it is on the same scale as the network data (\ie,
a one-second rolling sum of features). The model assumes the following PAI from
the server for each message: message text, the user who posted the message, and
the time stamp of when the message was first seen. The pseudocode for this step can be found in Algorithm~\ref{algo:log}.

\subsection{Deobfuscation Model} 

The output POL time-series for each IP's network traffic is then compared to
the POL of the user's messages. There are many options for similarity functions
at this stage. The results presented use the normalized cross-correlated (NCC)
as the similarity function. Equation~\ref{eq:crosscor} defines NCC where
$\overline{f(x)}$ is the complex conjugate of $f(x)$. Cross-correlation is
chosen as it is designed to detect when two time series vary similarly over
time. This captures if there is a certain IP address whose traffic elsewhere in
the network is usually seen before activity from an unknown user is seen. 

Normalized cross-correlation between time series is done by taking the time
series of an unknown user on the server and iteratively going through every
time series from possible source IP addresses. First, the source address is
filtered down to only include the same time range as the unknown user's
activity (a small amount of buffer room can be added here to account for
network delay). This is done to remove noise on either end of the communication
of interest. The user may have been using another site prior that would include
unrelated traffic, or an incorrect user may have sent unrelated messages. This
minimizes both of these types of noise. Then, the normalized cross-correlation
is computed between the two-time series, and the best, highest value is
selected. The source IP with the best normalized cross-correlation is
labeled as the IP that generated the traffic. This method evaluates the
performance of the scopes by looking at which scopes went into a given time
series. The pseudocode for this step can be found in Algorithm~\ref{algo:link}.

\begin{equation}
   cross\_cor_{x,y}(\tau) = \frac{E[(x_t-\mu_x)\overline{(y_{t+\tau}-\mu_y)}]}{\sigma_x\sigma_y} 
    \label{eq:crosscor}
\end{equation}


This approach ranks its output time series, where those with the highest
normalized cross-correlation rank highest and indicate the highest likely
deobfuscated attribution (\eg, the IP where the user is most likely
communicating from). Additionally, the method can return a list of the $k$ most
likely IP addresses for an analyst to search through. This technique could then
be applied to all users of a group of collaborating users to determine their
origin. Those users could then be tracked from conversation to conversation and
site to site to increase confidence in the results and determine other venues
where conversations between the same users are occurring (\eg, creating a
social network of users of interest).

\section{Evaluation}\label{sec:eval}

\begin{figure}
\centering
\includegraphics[width=0.45\textwidth]{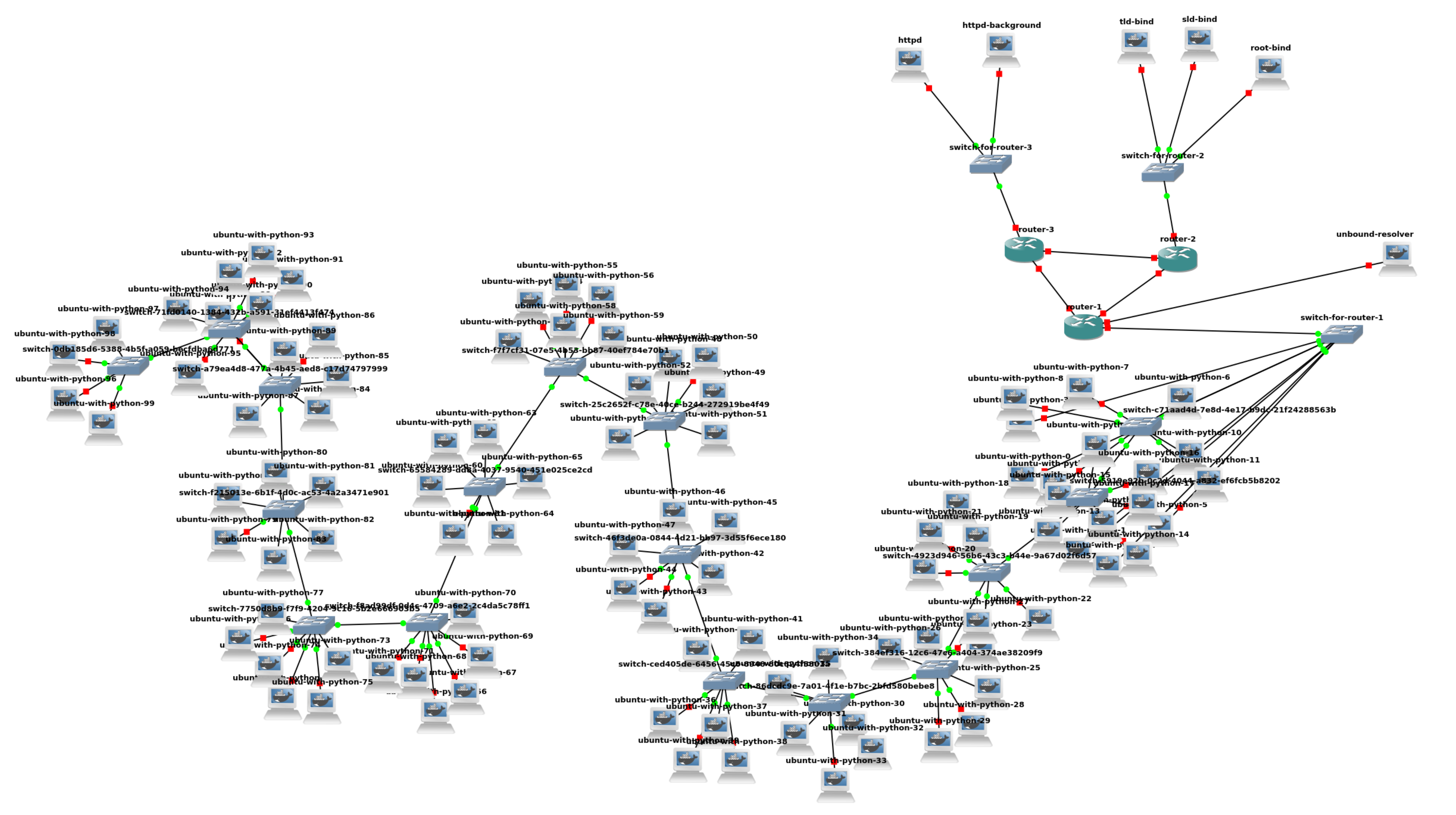}
\caption{The full 100-client GNS3 simulated topology.}
\label{fig:setup-without-vpn-100-clients}
\end{figure}

To evaluate our methodology, we used a multiyear-long message log of a well-known large-scale real-world social network application~\cite{subreddits}. 
The dataset consisted of 948,169 topic-driven interaction sites, was fully anonymized, and used timestamps to log users and conversational threads.  
To evaluate network behaviors we developed a simulation environment and recreated message postings with, and without, combinations of plain old DNS, DNS over HTTPS, DNS over TLS, and Virtual Private Networks.

\begin{figure}
  \centering
  \includegraphics[width=0.45\textwidth]{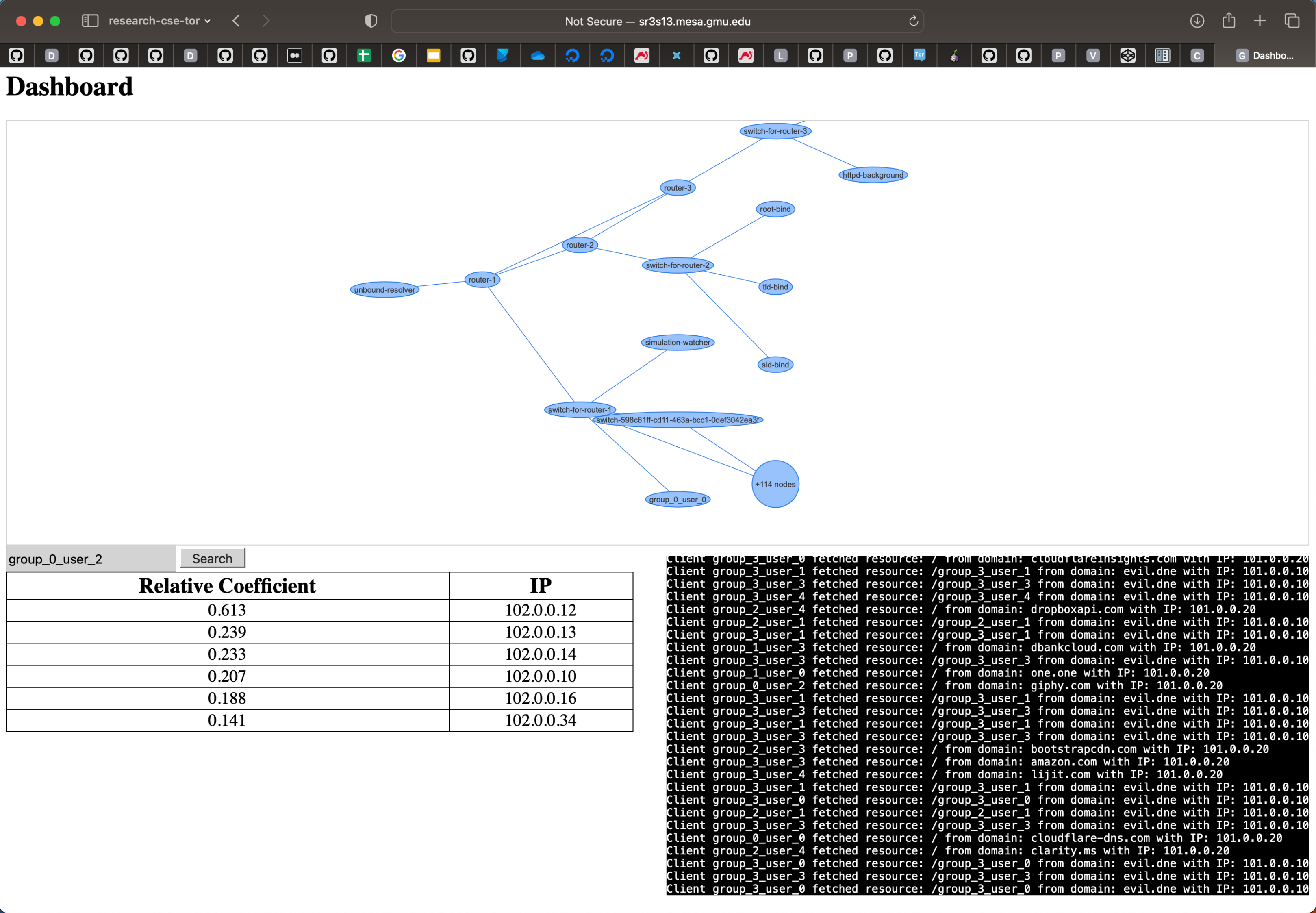}
  \caption{Proof-of-Concept (PoC) cybersecurity analyst dashboard}
  \label{fig:dashboard}
\end{figure}

Our full simulations involved 100 clients replaying random threads from the
social network site, depicted in
Figure~\ref{fig:setup-without-vpn-100-clients}. 
We decomposed network traffic (\eg, which scope traffic came from, packet headers, interarrival times, TCP segment information, and more) into \emph{features} that our methodology consumed.
After an initial grid search of
all available features, we examined one and two-feature combinations. We then
computed accuracy (absolute identification of identity) and recall@k (likelihood
the correct answer is in the top k results). We used this approach to create a
functional cybersecurity analyst dashboard, with a search-engine-like interface
that can indicate most likely deobfuscated sources for personas, ranked and
sorted by NCC, as seen in Figure~\ref{fig:dashboard}.

\begin{figure*}
\centering
\includegraphics[width=0.95\textwidth]{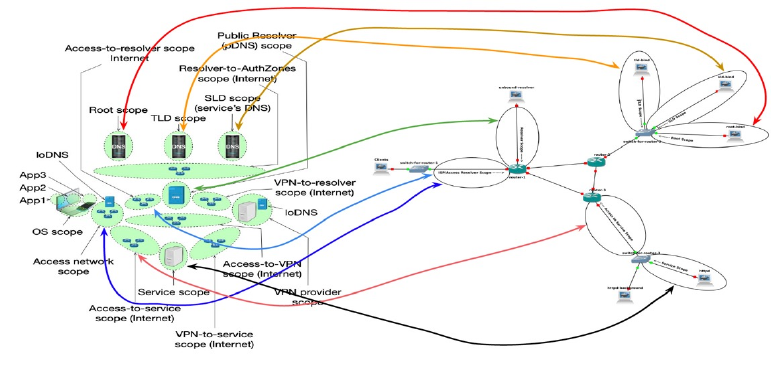}
\caption{This figure maps how we project our logical scopes topology into a GNS3 topology.}
\label{fig:scopes-gns3-mapping}
\end{figure*}

Our methodology relies on being able to measure data and meta-data that result
from combinations and interactions between a variety of network protocols. To evaluate the ability to perform deobfuscation, we generated network traffic
using NLnet Labs' unbound DNS resolver v1.13.1 (running plain old DNS, DNS over TLS, and DNS over HTTPS), ISC Bind servers v9.18.1 running as the Root, TLD and an SLD within a GNS3 topology that we designed to map our network scope
methodology into explicit routing and transport services, illustrated in
Figure~\ref{fig:scopes-gns3-mapping}.

In our simulations, we limited the size of our groups of interest to 
a representative group size of five members, as indicated by the literature~\cite{brodka2013ged}.
We selected random engaging conversations from our
dataset where multiple users posted multiple times. We defined engaging as a
conversation were multiple users posted more than five times each. We then
selected the five most talkative users from these conversations. This process
was repeated until we reached our desired number of groups. Our evaluations
simulated 20 groups, including actual message board conversations and unrelated
background traffic (randomly going to other sites every 15-30 seconds). 
%
%
To demonstrate the utility of this approach, our methodology is designed to
first, ingest network packet traces from the simulation as PCAP files and
convert network traffic into per-IP address data frames. When each IP address
is seen in traffic or is believed to exist in a DNS cache (estimated with a time window), our methodology adds
a time-based entry into that IP address' corresponding dataframe.
Next, a 1-second sliding sum is created over features extracted from network
traces in each scope where measurements exist. These features include IP packet
length and per-flow inter-packet timing.
After applying this windowing, we perform feature selection across scopes.
%
%
We then repeat these steps for features from the message board/service.
Then, in one of the most central steps in our methodology, we compare the PoLs
of measured features against the PoLs seen by message exchanges on the
message board. This comparison allows us to cross-correlate candidate
sources with the observed PoL on the actual network service (\ie the message
board).
Our final step is to evaluate how accurately our highest-correlated result
correctly identifies the actual source.
Among the benefits of this approach is an empirical spectrum of where
measurements can be taken across the Internet to provide the best accuracy and
recall@k for deobfuscation.

\subsection{ISP Selection}

Our simulations simulated 10 ISPs with clients uniformly distributed across
them. Our model showed that the performance at an ISP was equal to the
percentage of clients on that ISP multiplied by the performance of an ISP in
general. When tracking down a specific user, this equated to finding the user
with the metrics provided for a given experimental setup if the user's ISP was
monitored and being unable to find the user if the incorrect ISP was monitored.

\subsection{Results Layout}

The results of the experiments completed are summarized in the following graphs
and accompanying text. The results for a given PPT are shown for the
best-performing setup for a given scope across all metrics. For example, the
Service scope shows the results for the best-performing model that only uses
data from the Service scope. If more than one scope is used, the two scope
names are separated with a hyphen. ISPs in the experiments were labeled ISP1,
ISP2, ..., ISP10. If two scopes have equal accuracy, their order has no
significance. If a scope performs with all metrics 0 for a given PPT setup, the
model could not make any prediction when only given the scope. This could be
true for different reasons, including the scope being absent in the experiment
or no user-origin IP was ever seen in that scope (the model will only be able to evaluate the
IP which it has seen in the dataset). The global scope in the graphs refers to
performance when the model has access to every scope in the topology.

\subsection{Comparison of PPTs}

\begin{figure}
\centering
\includegraphics[width=0.45\textwidth]{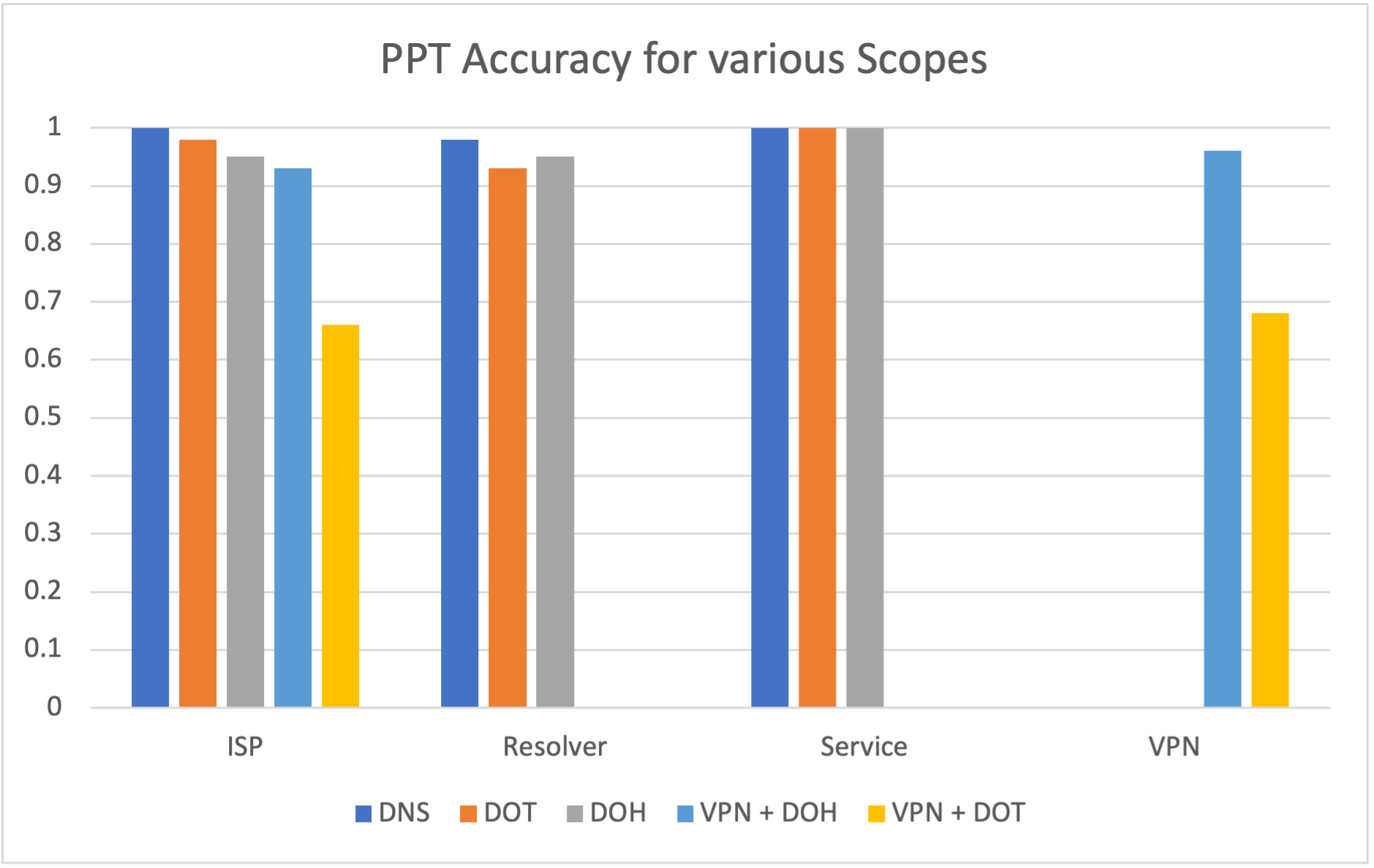}
\caption{Comparison of deobfuscation accuracy for various scopes}
\label{fig:comparePPT}
\end{figure}

Some PPT setups have different scopes available. 
When DoT or DoH were used with a VPN (\ie, DoT+VPN and DoH+VPN), the origin IP cannot be seen at the resolver or service, and thus, the model performed with 0 accuracy for those scopes. 
By contrast, when DNS, DoT, and DoH were used with no VPN, those PPTs have no metrics for the VPN scope.

The model presented achieved above 90\% accuracy for most PPT
combinations at multiple scopes as seen in Figure~\ref{fig:comparePPT}. DoH and DoT do provide
additional privacy-preserving capabilities over plain old DNS at the ISP, Resolver, or Service scopes. 
Our approach's strong ability to deobfuscate sources indicates that each these PPTs do not appear to provide enough privacy to be considered a complete solution. 
With the addition of a VPN, there is a noticeable shift in which scopes
can be monitored, as now some scopes which previously performed very well cannot
be used at all. 

\subsection{Baseline}

Initially, to evaluate our model, we ran a simulation where no clients were
using any PPT, and our model had complete knowledge of the communications. It
found that it was able to identify personas with 98\% and up when monitoring
from several different scopes. When the service, access-to-service, client
resolver or client ISP are monitored the model achieved a perfect score.
Although some of these best features may appear intriguing, it must be recalled that
as part of our data preprocessing, we were able to remove irrelevant data in a given
scope and only monitored DNS requests to the server, because (without a PPT) DNS query names were visible. As a result of this
preprocessing, some variables always return that same value for every packet.
This results in the features providing the same information with a scalar
difference, which is removed due to the normalization in our cross-correlation
calculation, resulting in those features being similar to a count of relevant
packets.

\subsection{DNS PPT}


The baseline simulations and analysis are repeated but with the addition of all
clients using DoH. The model results indicate how well the model can
deobfuscate a source address using specific features and scopes. The model can
identify the correct user with the same performance as when no PPT was used.
The only changes of note is that performance decreased by 3\% to 95\% at the
client's resolver and by 4\% to 95\% at the client's ISP. This shows 
that our methodology overcomes DoH's protections 95\% of the time.
No decryption is needed to gain these
results, and the model shows that even monitoring the number of packets over
time to the resolver reveals enough pattern of life to perform to deobfuscate
the client with high accuracy.


When running the simulation and analysis with DoT, the model results show a
similar level of privacy protection as with DoH: 
our approach is highly effective at deobfuscation.
These results show the same monitoring limitations and flexibility observed for DoH,
including no need to decrypt communications. We can see that, while DoT provides
almost the same level of privacy as DoH, more is needed in a comprehensive
privacy solution. 

When operationalizing our methodology in our dashboard,
recall@k is one of the most important measures for this type of model. This is
because it quantifies the experience and trust an analyst can have in such a
tool. A high recall@k tells the analyst the likelihood that the correct answer
is within the top k returned results allowing them to have an understanding for
the ``search'' performance of the model. 

When using multivariate time series analysis with this dataset we see that
there is varying impact when adding a second feature. For some stronger scopes,
our top performing scopes, there is little to no change. On the other hand with
some weaker scopes there can be a fairly significant drop in performance when
adding unhelpful features. Performance can drop by \~30\% accuracy in some
cases. This shows that there is still a need for an experienced analyst to be
able to get the most out of the model. We have seen that can sometimes occur
that two features when combined produce results that our lower than both of the
input features individual performance.

\subsection{VPN with DoT and DoH}

\begin{figure}
\centering
\includegraphics[width=0.45\textwidth]{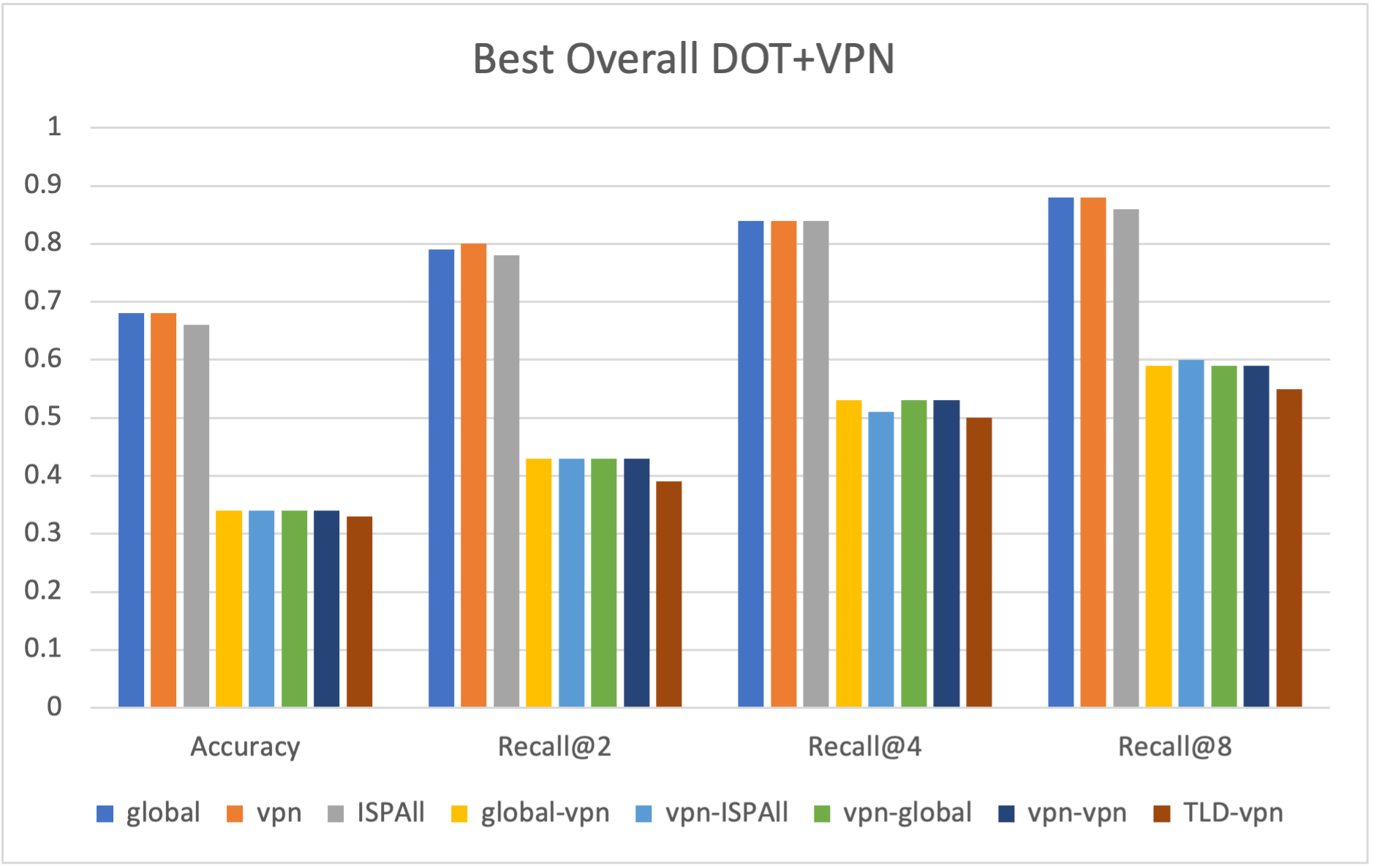}
\caption{ Highest performing deobfuscation in scopes when using DoT and VPN }
\label{fig:DoTVPNBest}
\end{figure}

\begin{figure}
\centering
\includegraphics[width=0.45\textwidth]{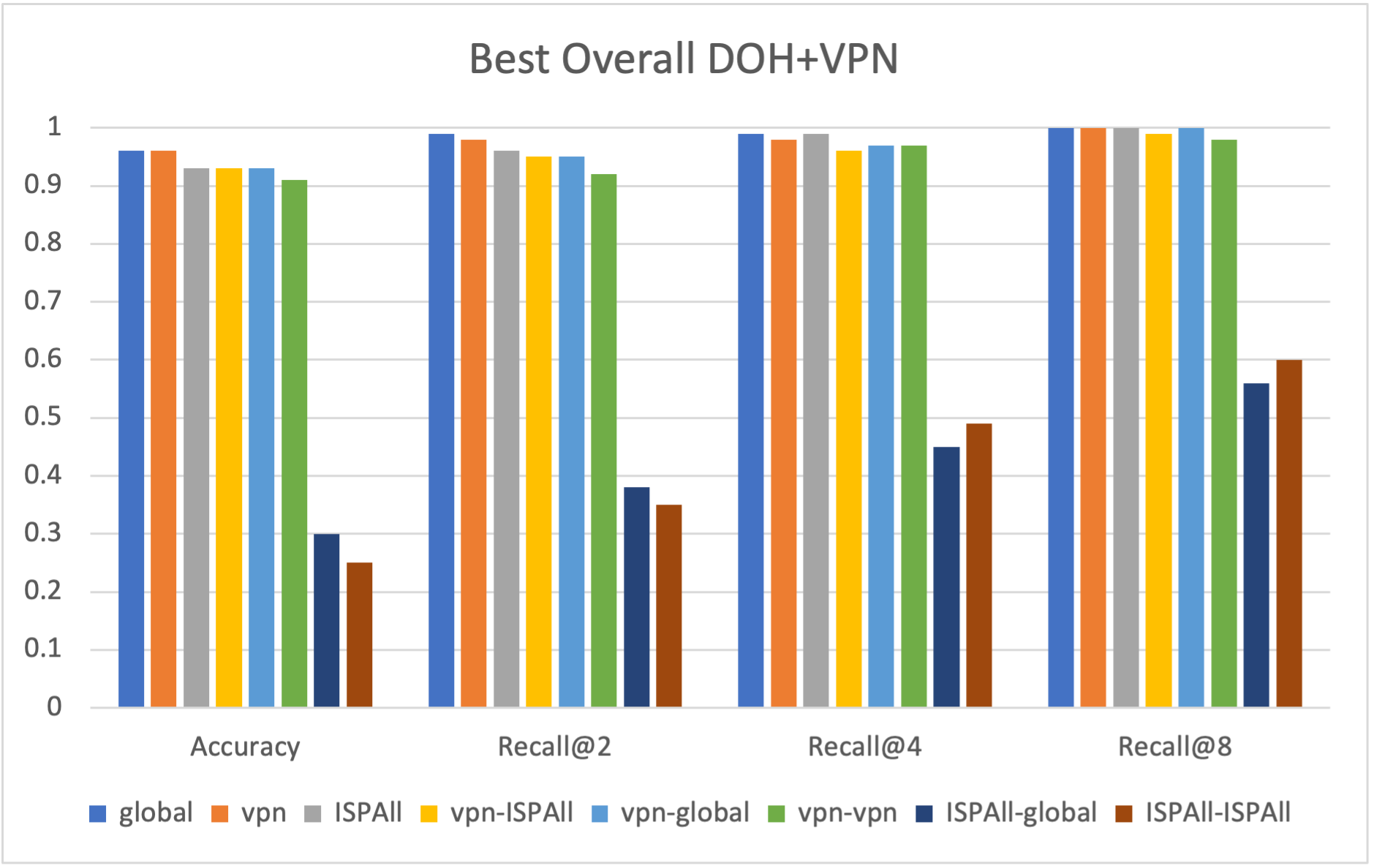}
\caption{ Highest performing deobfuscation in scopes when using DoH and VPN }
\label{fig:DoHVPNBest}
\end{figure}


When the client uses a VPN for all of their traffic (\ie sending both DNS and HTTPS traffic through the VPN), some scopes that previously performed well no
longer produce strong results. Monitoring at the client's resolver, the DNS root, TLD, SLD, access-to-service,
and server now can not deobfuscate client on their own as they cannot see the
client's original IP but only the VPN's IP. As a result, these scopes now show
as 0 for all metrics. 

While Figure~\ref{fig:DoHVPNBest} shows DoH with VPN is slightly more privacy-preserving than DoH, our methodology can still deobfuscate with roughly 90\% accuracy. 
%
%
On the other hand, Figure~\ref{fig:DoTVPNBest} shows DoT with VPN provides significantly more
privacy-preserving capabilities than any other method we tested. Accuracy drops
by around 28\%-33\%, and recall@k follows a similar pattern. This is interesting as DoT and DoH are believed to provide the same privacy and security
protections as they both use TLS. 

In our experiments, we have seen that subnet information is leaked when using a
VPN. This is due to the service employing techniques best to serve the clients
with a better CDN location, specifically EDNS client subnet zero~\cite{rfc7871}. This information could be
used in a multistage analysis of our existing method, where the leaked subnet
information determines the client's ISP or VPN. This would be followed by
tapping the given scope, which would allow for deobfuscation of the client.
Figure~\ref{fig:DoTVPNBest} and Figure~\ref{fig:DoHVPNBest} show the best possible performance of our methodology, with full visibility of all Internet traffic.

\subsection{MultiScope Performance}

There is an important takeaway when applying our method for using multiple
features on the data set provided. 
Traffic cannot always be observed from an optimal scope. 
For example, traffic from a user's resolver may not be available.
However, using multivariate analyses over \emph{multiple} scopes together can produce equivalent results.
We have seen that the method presented can
be used to transform multivariate time series of network data into univariate
time series using TDA that can capture the pattern of life of users online.
Although the current method does not provide results that are strictly better
than the univariate version of the model, it can be seen that the
transformation using TDA often does not lose any resolving accuracy while
taking advantage of multiple features from different scopes. Table
~\ref{tab:tdaDrop} shows the best multivariate performance compared to the best
univariate performance for different PPT. This comparability is an important finding as this
is not an obvious outcome from using the techniques. TDA inherently transforms
the data into a smaller space; thus, the ability to do so while still
preserving the key features of the pattern of life is very promising future work. 

\begin{table}
    \begin{center}
        \begin{tabular}[c]{|l|l|l|}
            \hline
            {\bf PPT} & {\bf No-TDA} & {\bf TDA}\\ \hline
            \hline
            DoH & 100\% & 100\%\\
            DoT & 100\% & 99\%\\
            DoH+VPN & 96\% & 93\%\\
            DoT+VPN & 68\% & 34\%\\
            \hline
        \end{tabular}
    \end{center}
    \caption{ Preservation of POL when TDA is used }\label{tab:tdaDrop}
\end{table}





\subsection{The Playbook}

To operationalize our methodology,
the first step is to scrape the message board in question. Specifically, for
the user in question, scrape as many messages as possible (across all threads
the user was using on the site if possible) and the time each message was
posted (as much resolution on the time stamp as possible).
Our methodology then looks for
the pattern of life (POL) of these messages across monitored traffic to determine
the origin IP of the user. The best monitoring places depend on the user's
Privacy Preserving Technology (PPT). Some scopes always work to find the user: such as the
user's origin ISP. Though, clearly, one cannot always know the user's origin ISP without
some guilty knowledge. One could monitor every ISP, but this is not realistic.

Some scopes give general information, which can be a helpful place to
start. The access-to-service or resolver scopes would be ideal as we can quickly determine
if an IP's POL matches our user (\eg an origin IP or a VPN server). If there is
no precise match, then our user is likely using some PPT, and as a result, we
need to move our monitoring closer to the user's source.  

Suppose we know that the user did not use any PPTs. In that case, we can
monitor where the HTTPS traffic passed through (\eg origin ISP or
access-to-service) or anywhere the complete DNS went (i.e., resolver. Note:
root DNS, TLD, and SLD only have a partial view of the DNS traffic due to
caching at the resolver. If the client also caches DNS, then the resolver would
also not have a complete view of the DNS). Monitoring at any of these scopes is
very similar, so tapping them is equal.

If the user only uses a DNS PPT, we can find them using the same techniques as if
no PPT were used. We have found only a minor degradation in performance when
adding a DNS PPT.

If the user uses a VPN, the goal should be to find a scope pre-VPN or at the
VPN. These can be used to find the origin IP. Unfortunately, this is a much
larger space with many more scopes. If the VPN or user's ISP identified a
different way then this would become  a much easier problem. Once a pre-VPN
scope is found it is reasonable to be able to find the user if they are using
DoH and harder but still possible if they are using DoT.



\section{Future Work}\label{sec:future}

{\bf MultiScope Performance:}
There is still work to be done to determine how to best normalize and weight
features from different scopes prior to the transformation in order to gain the
most benefits from using TDA in this manner. There also may be a benefit in
combining features originating from the same scope and different scopes in
different ways.

It is additionally interesting that the largest performance drop with TDA was
also the most challenging to deobfuscate when the univariate model was used
PPT. Performance with DoT and VPN dropped by 34\%, half the performance when
the univariate model was used. The reason for this drop being larger than the
others is unknown and should be investigated further as it is not known to
attribute this change to a success of DoT and VPN or a shortcoming of the
model.

{\bf Dashboard:}
The cybersecurity dashboard provides a general interface to using our methodology to deobfuscate sources.
Our future work will evolve and enhance its basic utility to augment its current focus, and include more operational cybersecurity analyst use-cases.

\section{Related Work}\label{sec:relwork}

This work exists in the broader domain of the classification of encrypted
communications. Machine learning approaches have dominated this field in
recent years due to their ability to overcome noisy data and model complex
problems. Most work in this field has focused on tabular and flow summary
statistics. Important work has been done using these techniques in botnet
detection~\cite{Alauthman20}, traffic classification~\cite{Alshammari09}, VPN
detection~\cite{didarknet}~\cite{Gil2016}, and tor
deobfuscation~\cite{darkdetect}~\cite{Lashkari2017}. Although these techniques
can be effective at some classes of problems, they leave some performance on the
table by transforming data that is inherently a time series into summary
statistics. In addition, these techniques use an overly simplified
representation of the Internet. These techniques, at large, focus on the inputs and
outputs of a simplified system when, in reality, if one of these models were to
be deployed, one would need to decide where to capture the traffic on the
Internet. It is not a given that all places on the Internet would provide the
same level of performance. This work addresses this concern to aid the
analyst in knowing the best places to drop monitors across the Internet,
given the threat model and PPT deployed. Additionally, it is an open problem for best-combining information from different scopes across the Internet once they are gathered. 

Another related field is that of author attribution and account linking. This
field is concerned with determining if two personas online belong to the same
real-world individual utilizing text
analysis~\cite{noa2021}~\cite{Silvestri2015LinkingAA}. This problem has many
similarities to the goal of our work, but our work is focused on network patterns of
life instead of natural language processing techniques. This field has access
to much larger text datasets and thus can take advantage of large transformer
models. Unfortunately, such an open dataset does not exist for the work covered
in this paper, so simulations of a much smaller size in terms of users and number/length of conversation must take
place. This makes deep learning models out of the question due to the limited
size of the datasets that can be reasonably simulated. Our work has shown that
accessing the message text is sometimes unnecessary to link a user to
their IP address. 

The techniques used for time series analysis with TDA have shown success in
other complex multivariate domains~\cite{seversky16}. TDA has shown success in
modeling gene expression~\cite{perea15}, cryptocurrency scam
analysis~\cite{BitcoinHeist}, network anomaly
detection~\cite{bruillard2016anomaly}, passive IOT
classification~\cite{collins2020passive}, and network activity
prediction~\cite{GABDRAKHMANOVA2019616}. Persistence
landscapes~\cite{bubenik15b}~\cite{bubenik15a} specifically have shown success
in financial market analysis~\cite{gidea17}, cryptocurrency
prediction~\cite{gidea18},  internet of things device
classification~\cite{TDA_IOT}, deep learning layers~\cite{topolayer2020}, and
medical applications~\cite{bendich2014persistent}. These works have shown that
TDA and persistence landscapes are not only capable of performing
competitively for machine learning and time series analysis tasks but can
outperform other methods. The topological features captured through TDA are
challenging for classical statistics models to capture. It is for these reasons 
TDA was the tool chosen for this work.

\section{Conclusion}\label{sec:conc}

In this paper, we illustrate the effectiveness of using our novel methodology
to successfully deobfuscate network sources who are using Privacy Preserving
Technologies (PPTs) to hide their source addresses while conducting malfeasance
on public forums. Using a deidentified multiyear-long message log from a large scale
social network site, we illustrated that a Pattern of Life (PoL) can readily be
constructed and used to correctly deobfuscate source addresses, even when DNS
over TLS (DoT), DNS over HTTPS (DoH), and/or Virtual Private Networks (VPNs)
were used. We found that our proof of concept cybersecurity dashboard was able
to correctly identify obfuscated sources with between 0.9 to 1.0 accuracy (out
of 1.0), depending on which PPTs were in use and where observations were made.

\bibliographystyle{IEEEtran}
\bibliography{paper,rfc}

\appendix
\section{Algorithm Pseudocode}\label{app:algo}

Data wrangling, Algorithm~\ref{algo:data}, for the network aims to split up packets based on their
importance to the IPs seen in the dataset. This step finishes with transforming
the data into a time series and applying the TDA transformations.

\begin{algorithm}
\caption{Data Wrangling: $O(n^2)$ }\label{algo:data}
\begin{algorithmic}[1]
\Require $S$, a list of scopes (a scope is a list of PCAPs)
\Require $w$, window to use for TTL approximation
\Function{DataProccessing}{}
    \State $output \gets []$
    \ForAll {$scope \in S$}
        \ForAll {$pcap \in scope$}
            \ForAll {$packet \in pcap$}
                \State $append(output[packet.srcIP], packet)$
                \State $append(output[packet.dstIP], packet)$
                \ForAll {$ip \in output$}
                    \If {$overlap(output[ip], packet.time, w)$}
                    \State $append(output[ip], packet)$
                    \EndIf
                \EndFor
            \EndFor
        \EndFor
    \EndFor
    \ForAll{$ip \in output$}
        \State downselect features for $output[ip]$
        \State $output[ip] \gets rollingSum(output[ip])$
        \State $output[ip] \gets TDA\-PL(output[ip])$
    \EndFor
    \State \textbf{return} $output$ \Comment{A mulitvariate time series for every IP address}
\EndFunction
\end{algorithmic}
\end{algorithm}

The data transformations for the scraped server messages, Algorithm~\ref{algo:log}, follows
a very similar pattern as the data wrangling of the network traces but only
groups messages from users who are known to be the same together.

\begin{algorithm}
\caption{Server Log Prep: $O(n^2)$ }\label{algo:log}
\begin{algorithmic}[1]
\Require $F$, a list of message board log files
\Function{LogProccessing}{}
    \State $output \gets []$
    \ForAll {$file \in F$}
        \ForAll {$message\in file$}
            \State $append(ouput[message.user], message)$
        \EndFor
    \EndFor
    \ForAll{$user \in output$}
        \State $output[user] \gets rollingSum(output[user])$
        \State $output[ip] \gets TDA\-PL(output[ip])$
    \EndFor
    \State \textbf{return} $output$ \Comment{A mulitvariate time series for every username}
\EndFunction
\end{algorithmic}
\end{algorithm}

The deobfuscation model, Algorithm~\ref{algo:link}, aims to compare the server logs with
the network traces to find the same PoL between IPs and users. Its similarity
function could be replaced but our work used NCC, Algorithm~\ref{algo:sim}, and
TDA, Algorithm~\ref{algo:tda}, to determine time series similarity.

\begin{algorithm}
\caption{Deobfuscation Model: $O(n^4)$}\label{algo:link}
\begin{algorithmic}[1]
\Require $I$, POL of each IP (Algorithm~\ref{algo:data})
\Require $U$, POL of each user (Algorithm~\ref{algo:log})
\Function{DeobfuscationModel}{}
    \State $output \gets []$
    \ForAll{$user \in U$}
        \State $bestIP \gets 0.0.0.0$
        \State $best \gets 0$
        \State $ranking \gets []$
        \ForAll{$ip \in I$}
            \State $s \gets similarity(ip, user)$
            \If $s > best$
                \State $best \gets s$
                \State $bestIP \gets ip$
            \EndIf
            \State $ranking.push([s, ip])$
        \EndFor
        \State $output[user] \gets [bestIP, sort(ranking)]$
    \EndFor
    \State \textbf{return} $output$ \Comment{An ordered list of IPs for each username ordered by similarity}
\EndFunction
\end{algorithmic}
\end{algorithm}

\begin{algorithm}
\caption{Time-series Similarity: $O(n^2)$ }\label{algo:sim}
\begin{algorithmic}[1]
\Require $a, b$, time series to compare
\Function{similarity}{}
    \State $b \gets copyRange(b, a)$
    \State \textbf{return} $NCC(a, b)$ \Comment{The similarity from 0 to 1}
\EndFunction
\end{algorithmic}
\end{algorithm}

\begin{algorithm}
\caption{TDA\-PL: $O(n^2)$ }\label{algo:tda}
\begin{algorithmic}[1]
\Require $t$, a time series
\Require $w$, window size
\Require $d$, dimension of homology to use
\Require $m$, max filtration distance
\Require $l$, number of landscapes to compute
\Function{TDA\-PL}{}
    \State $output \gets []$
    \State $windows \gets split(t, w)$
    \For{$i \gets 0,length(windows)$}
        \State $ph \gets ripsFiltration(windows[i], k, m)$  \Comment $O(n^2)$
        \State $pl \gets persistenceLandscape(ph, l)$  \Comment $O(n\log(n))$
        \State $l2 \gets L^2(pl)$  \Comment $O(n)$
        \State $output[i] \gets l2$
    \EndFor
    \State \textbf{return} $output$ \Comment{A univariate time series}
\EndFunction
\end{algorithmic}
\end{algorithm}

\end{document}